\pdfoutput=1

\documentclass[12pt,nofootinbib]{article}
\usepackage{amsmath}
\usepackage{amssymb}
\usepackage{graphicx}

\numberwithin{equation}{section}

\def\aD3{$\overline{\mathrm{D}3}$}

\begin{document}

\begin{titlepage}

\setcounter{page}{1} \baselineskip=15.5pt \thispagestyle{empty}

\begin{flushright}
{\footnotesize ICRR-Report-546}\\
{\footnotesize IPMU09-0064}\\
{\footnotesize UTAP-613}\\
{\footnotesize RESCEU-12/09}
\end{flushright}
\vfil

\bigskip\
\begin{center}
{\LARGE  Curvatons in Warped Throats}
\vskip 15pt
\end{center}

\vspace{0.5cm}
\begin{center}
{\large Takeshi Kobayashi$^{\ast}$\footnote{takeshi.kobayashi@ipmu.jp} and
 Shinji Mukohyama$^{\dagger}$\footnote{shinji.mukohyama@ipmu.jp}}
\end{center}

\vspace{0.3cm}

\begin{center}
\textit{$^{\ast}$ Institute for Cosmic Ray Research, The University of
 Tokyo, \\ 5-1-5 Kashiwanoha, Kashiwa, Chiba 277-8582, Japan}\\

\vskip 4pt
\textit{$^{\dagger}$ Institute for the Physics and Mathematics of the
 Universe (IPMU), \\ The University of Tokyo, 5-1-5 Kashiwanoha,
 Kashiwa, Chiba 277-8582, Japan}
\end{center} \vfil

\vspace{0.8cm}

\noindent
We present a curvaton model from type IIB string theory compactified on
 a warped throat with approximate isometries. Considering an
 (anti-)D3-brane sitting at the throat tip as a prototype standard model 
 brane, we show that the brane's position in the isometry directions can
 play the role of curvatons. The basic picture is that
 the fluctuations of the (anti-)D3-brane in the angular isometry
 directions during inflation eventually turns into the primordial
 curvature perturbations, and subsequently the brane's oscillation
 excites other open string modes on the brane and reheat the
 universe. We find in the explicit case of the KS throat that a wide
 range of parameters allows a consistent curvaton scenario. It is also
 shown that the oscillations of branes at throat tips are capable of
 producing large non-Gaussianity, either through curvature or
 isocurvature perturbations. Since such setups naturally arise in 
 warped (multi-)throat compactifications and are constrained by
 observational data, the model can provide tests
 for compactification scenarios. This work gives an explicit example
 of string theory providing light fields for generating
 curvature perturbations. Such mechanisms free the inflaton from being
 responsible for the perturbations, thus open up new possibilities for
 inflation models.

\vfil

\end{titlepage}

\newpage
\tableofcontents

\section{Introduction}
\label{sec:intro}

The inflationary era in the early
universe~\cite{Starobinsky:1980te,Sato:1980yn,Guth:1980zm} not only sets 
appropriate initial conditions for the subsequent Hot Big Bang
cosmology, but also provides initial inhomogeneities of the universe and
seeds the formation of large scale structures. However, since inflation
likely takes place at energy scales way beyond our current
experimental reach, a detailed understanding of the origin of
inflationary cosmology has remained a major theoretical challenge. 
In order to connect the underlying high-energy physics and current
cosmological observations, it is essential to come up with explicit
models in a UV-complete framework. This has stimulated numerous attempts
to embed inflationary cosmology within string theory which is so far
our best candidate for a UV-completion of quantum field theory and
gravity. (For a review, see e.g.,
\cite{Cline:2006hu,Kallosh:2007ig,Burgess:2007pz,McAllister:2007bg,Baumann:2009ni}.) 

An important feature of string cosmology is its high sensitivity to the
physics of string compactification. This imposes stringent
restrictions on string inflationary models, and it is nontrivial whether
the inflaton can generate the primordial curvature perturbations. 
In order to avoid decompactification, low-scale inflation
is favored, which often leads to the amplitudes of the resulting
curvature perturbations being too small compared to the observed
COBE normalization value~\cite{Komatsu:2008hk}. Recent studies 
\cite{Badziak:2008yg,Abe:2008xu,Conlon:2008cj,Badziak:2008gv,Chen:2009nk}
have shown that the inflation scale can be disentangled from the 
low energy scale of supersymmetry phenomenology, but still the scale of
inflation is required to lie within the energy range compatible with stable
compactifications. 
Furthermore, moduli stabilization effects generally introduce steep
potentials for the inflaton, which was shown explicitly for the case of
warped 
D-brane inflation
models~\cite{Kachru:2003sx,Baumann:2006th,Baumann:2007np,Krause:2007jk,Baumann:2007ah}
in the form of the $\eta$-problem. Although there exist rapid-roll
inflationary attractors for such
potentials~\cite{Linde:2001ae,Kofman:2007tr,Kobayashi:2008rx,Kobayashi:2009nv},
the inflaton itself cannot produce a scale-invariant perturbation
spectrum. 

Thus, one can expect some field(s) other than the inflaton to have
generated the curvature perturbations, as in the case of the curvaton
scenario~\cite{Linde:1996gt,Enqvist:2001zp,Lyth:2001nq,Moroi:2001ct}. Realization
of such mechanisms can drastically relax the constraints on the inflaton
potential, setting inflation models free from the standard slow-roll
type~\cite{Dimopoulos:2002kt}.\footnote{Some earlier discussions on
curvatons in string theory appeared in \cite{Pilo:2004mg}. There are 
also other known mechanisms for generating curvature perturbations,
such as modulated reheating~\cite{Dvali:2003em,Kofman:2003nx} and the Lyth
effect~\cite{Lyth:2005qk,Alabidi:2006wa}. Stringy realizations of the
latter have been attempted in
\cite{Lyth:2006nx,Leblond:2006cc,Chen:2008ada}.}

In this paper, we present a simple curvaton model
from string theory compactified on a warped throat with approximate
isometries. A good example of such throats is the deformed
conifold~\cite{Klebanov:2000hb} in 
type IIB string theory, where it has been shown that fluxes and
nonperturbative effects can stabilize all its
moduli~\cite{Giddings:2001yu,Kachru:2003aw}. 
Considering an (anti-)D3-brane sitting at the throat tip as a prototype
standard model brane,\footnote{For an attempt to realize the standard
model on \aD3-branes at throat tips, see e.g. \cite{Cascales:2003wn}.}
we show that the brane's position in the isometry 
directions can play the role of the curvaton. The basic picture is that
the fluctuations of the (anti-)D3-brane in the angular isometry
directions during inflation eventually turns into the primordial
curvature perturbations, and subsequently the brane's oscillation
excites other open string modes on the brane and reheat our
universe.\footnote{Reheating the universe from D-brane oscillation was
also considered in \cite{Brodie:2003qv,Mukohyama:2007ig}, however, the
details differ from what we are going to discuss in this paper.} We find
in the explicit case of the Klebanov-Strassler (KS)
throat~\cite{Klebanov:2000hb} that a wide range of parameters allows a
consistent curvaton scenario. 

An important feature of the model is that the generated perturbation
spectrum can also have large non-Gaussianity. In this regard, later in
this paper, we also consider effects on the perturbation spectrum when
the internal manifold contains multiple throats with (anti-)D3-branes at
their tips. Even if the branes' fluctuations in the isometry directions
produce only a negligible amount of curvature perturbations, they turn
out to be capable of contributing substantially to the non-Gaussianity
either through adiabatic or isocurvature perturbations. These features
provide tests not only for the curvaton model, but also for multi-throat
compactifications. 

The paper is organized as follows. In Section~\ref{sec:curvaton} we lay
out the requirements for a general curvaton scenario to successfully
produce the primordial perturbations. In Section~\ref{sec:CWT} we
discuss the dynamics of an \aD3-brane at the tip of a warped deformed
conifold, and show that its oscillation can serve as a curvaton. Then in
Section~\ref{sec:perturbations} we explore the parameter space and study
several cases where the curvaton produces curvature perturbations and/or
non-Gaussianity. Finally in Section~\ref{sec:con} we conclude and give
discussions on future prospects.

\section{Requirements for the Curvaton}
\label{sec:curvaton}

A curvaton~\cite{Linde:1996gt,Enqvist:2001zp,Lyth:2001nq,Moroi:2001ct,Dimopoulos:2003ss}
is a light field, hence during inflation it acquires fluctuations that
are nearly scale invariant and Gaussian. Although its energy density
is initially negligible, after inflation the curvaton starts oscillating
around its potential minimum and behaves like nonrelativistic matter. As
its energy density grows relative to radiation, its fluctuations
increasingly contribute to the curvature perturbation until the curvaton
decays or dominates the universe. Let us begin by laying out the
requirements for a consistent curvaton scenario. 

We assume the inflation energy to be converted to radiation
right after inflation ends (a stringy realization of such a setup is
discussed in Section~\ref{sec:perturbations}), and also the curvaton to
be out of equilibrium until it decays into SM particles, or particles that
eventually turn into SM particles. We denote the curvaton mass as $m$,
its decay rate $\Gamma$, and its field value during inflation
$\sigma_*$. When the curvaton potential is quadratic, then its
energy density is negligible during inflation if
\begin{equation}
 \frac{m^2 \sigma_*^2}{V_{\mathrm{inf}}} \sim 
 \frac{m^2 \sigma_*^2}{M_{\mathrm{Pl}}^2 H_{\mathrm{inf}}^2} 
 \ll 1, \label{1}
\end{equation}
where $V_{\mathrm{inf}}$ and $H_{\mathrm{inf}}$ are respectively the
inflation energy scale and Hubble parameter, and $M_{\mathrm{Pl}}$ is the reduced
Planck mass. We also require the curvaton mass to be sufficiently
smaller than the Hubble parameter, 
\begin{equation}
 \frac{m^2}{H_{\mathrm{inf}}^2} \ll 1. \label{2}
\end{equation}
Setting the cosmic time $t$ by $\rho \sim M_{\mathrm{Pl}}^2 /t^2$, then inflation
ends at $t_{\mathrm{end}} \sim H_{\mathrm{inf}}^{-1}$. Due to the Hubble
damping, the curvaton remains at $\sigma_*$ until the Hubble parameter
drops below its mass at $t_{\mathrm{osc}}\sim m^{-1}$ and then the
curvaton starts to oscillate. If the curvaton's lifetime is long enough,
then it dominates the energy density of the universe at
time~$t_{\mathrm{dom}}$ satisfying
$V_{\mathrm{inf}} (t_{\mathrm{end}}/t_{\mathrm{dom}})^2 \sim m^2
\sigma_*^2 (t_{\mathrm{osc}}/t_{\mathrm{dom}})^{3/2}$, which gives
\begin{equation}
 t_{\mathrm{dom}} \sim \frac{M_{\mathrm{Pl}}^4}{m \sigma_*^4}.
\end{equation}
This will be after the onset of the curvaton oscillation if and only if
\begin{equation}
 \frac{\sigma_*}{M_{\mathrm{Pl}}}< 1 \label{3}
\end{equation}
is satisfied. Later in (\ref{trivial3}) we will see that the
condition~(\ref{3}) is automatically satisfied in our model. It should
also be noted that (\ref{1}) is trivially satisfied under (\ref{2}) and
(\ref{3}). 

We further require the curvaton decay at $t_{\mathrm{dec}} \sim
\Gamma^{-1}$ to be after $t_{\mathrm{osc}}$, 
hence we impose
\begin{equation}
 \frac{\Gamma}{m} < 1. \label{4}
\end{equation}
This requirement turns out to be inevitable for our model, when we
introduce additional microscopic constraints in 
Subsection~\ref{subsec:con} (cf. (\ref{trivial4})).
If the curvaton contributes a significant fraction of
the total energy density, then it should decay before the Big Bang
Nucleosynthesis (BBN) at $\rho_{\mathrm{BBN}}\sim (1\, \mathrm{MeV})^4
\sim 10^{-84} M_{\mathrm{Pl}}^4$, which requires 
\begin{equation}
 10^{42} \frac{\Gamma}{M_{\mathrm{Pl}}} > 1. \label{5}
\end{equation}

The curvature perturbation produced by the curvaton stops growing when
the curvaton decays or dominates the universe. For the generated
perturbation to be nearly Gaussian, the field perturbation of the curvaton
should be smaller than the unperturbed value,
\begin{equation}
 \frac{H_{\mathrm{inf}}}{\sigma_*} < 1. \label{6}
\end{equation}
This requirement is also inevitable for our model, as we will see in
(\ref{bprime}). Then, the final curvature perturbation is given by 
\begin{equation}
 \zeta \sim r \frac{\delta \rho_{\sigma}}{\rho_{\sigma}} \sim
 r \left[2 \frac{\delta\sigma_*}{\sigma_* } + \left(\frac{\delta
		      \sigma_*}{\sigma_*}\right)^2  \right].
 \label{czeta}
\end{equation}
Here, $r$ is the energy density fraction of the curvaton
at the decay epoch, which one can estimate as
\begin{equation}
 r \equiv \left. \frac{\rho_{\sigma}}{\rho} \right|_{\mathrm{dec}} \sim
 \mathrm{min} \left\{\left(
	       \frac{t_{\mathrm{dec}}}{t_{\mathrm{dom}}}\right)^{1/2},\
	       1\right\}   \sim
 \mathrm{min} \left\{ \frac{m^{1/2} \sigma_*^2}{\Gamma^{1/2}
		       M_{\mathrm{Pl}}^2},\ 1\right\} .
\end{equation}
Therefore the power spectrum of the generated curvature perturbation is
\begin{equation}
 \mathcal{P}_{\zeta}^{1/2} \sim r \frac{H_{\mathrm{inf}}}{\sigma_*}
  \lesssim \mathcal{P}_{\mathrm{COBE}}^{1/2} \sim 10^{-5}, \label{7.5}
\end{equation}
where the upper bound on the amplitude is set by the COBE
normalization~\cite{Komatsu:2008hk}. (If the curvaton is the only
source for the perturbations, then the inequality should be saturated.) 
The second term in the far right hand side of (\ref{czeta}) contributes
to non-Gaussian perturbations. In terms of the non-linearity
parameter~$f_{\mathrm{NL}}$~\cite{Komatsu:2001rj}, the non-Gaussianity
in the curvature perturbation\footnote{We discuss non-Gaussianity
produced through isocurvature perturbation in
Subsection~\ref{subsec:nG-iso}.} can be estimated as\footnote{We
have assumed that the fluctuation of the curvaton is not correlated with
that of any other field which generates curvature perturbations.}
\begin{equation}
 f_{\mathrm{NL}} \sim \frac{r^3}{\mathcal{P}_{\mathrm{COBE}}^{2}} \left(
			   \frac{H_{\mathrm{inf}}}{\sigma_*}\right)^4 
  \lesssim 100 ,
 \label{8}
\end{equation}
where we have also presented the observational upper bound from
WMAP~\cite{Komatsu:2008hk}. It is clear from (\ref{7.5}) and (\ref{8})
that if the curvaton dominates the universe before decay, i.e. $r\sim
1$, then the produced $f_{\mathrm{NL}}$ is at most of order one.

\section{Curvatons in Warped Throats}
\label{sec:CWT}

We now show an explicit realization of the curvaton in the context of
warped type IIB compactifications. We consider the six-dimensional
internal space to be compactified to a (conformally) Calabi-Yau (CY)
space which includes warped deformed conifold throat regions. Assuming
that we are living on an \aD3-brane sitting at the tip of the conifold
throat, the angular position of the \aD3 in the isometry
directions plays the role of curvatons. (The reason why we choose an \aD3
and not a D3 will be given at the end of Subsection~\ref{subsec:WDC}.)

However, since the throat is glued to the bulk CY which in general do
not preserve such isometries, the throat isometries are broken. Also,
nonperturbative effects that stabilize the K\"ahler moduli confine the
\aD3 to certain loci on the tip. The warping of the throat suppresses
all such effects at the tip, and consequently the angular position of
the \aD3 receives small mass and couplings to other open
string modes on the brane.

\subsection{Warped Deformed Conifold}
\label{subsec:WDC}

Let us begin by discussing the geometry of the warped deformed
conifold. Away from the tip, the deformed
conifold~\cite{Candelas:1989js} is well approximated by
a cone whose five-dimensional base space is $S^2 \times S^3$, with the
isometry group $SU(2) \times SU(2) \times U(1)$. As one approaches the
tip, the $S^2$ shrinks to zero size, while the $S^3$ remains finite. 

The supergravity solution for the warped deformed conifold was
found by KS~\cite{Klebanov:2000hb}, and is also included in the class of
flux compactifications considered by Giddings, Kachru, and Polchinski
(GKP)~\cite{Giddings:2001yu}. There it was shown that the KS solution
can be constructed by turning on $M$ units of R-R flux $F_3$ on the
$S^3$ at the tip, and also $-K$ units of NS-NS flux $H_3$ on the dual
cycle,
\begin{equation}
 \frac{1}{2\pi \alpha'} \int_A F_3 = 2\pi M, \qquad 
 \frac{1}{2\pi \alpha'} \int_B H_3 = -2\pi K.  \label{MK}
\end{equation}
The product of $M$ and $K$ produces the background D3-charge~$N$. By
lifting the KS solution to F-theory, the tadpole-cancellation condition
relates $N$ to the Euler number $\chi$ of the corresponding CY
four-fold. The known maximal value for the Euler number $\chi_* =
1820448$~\cite{Klemm:1996ts} 
implies an upper bound on the background charge number,
\begin{equation}
 N \equiv MK \leq \frac{\chi_*}{24} \sim 10^5. \label{euler}
\end{equation}
It should also be noted that the supergravity description is valid for
$g_s M \gg 1$, where $g_s$ is the string coupling. 

We write down the leading order background geometry in the following
form, 
\begin{equation}
 ds^2 = h(r)^2 g^{(4)}_{\mu\nu} dx^{\mu} dx^{\nu} + h(r)^{-2} 
\left( dr^2 + r^2 g^{(5)}_{mn} d\theta^m d\theta^n\right),
\end{equation}
where $x^{\mu}$ ($\mu = 0,1,2,3$) are the external four-dimensional
coordinates, $r$ is the radial coordinate which decreases as one
approaches the tip of the throat, and $\theta^m$ ($m=5,6,7,8,9$) are the
five-dimensional angular coordinates. $h(r)$ is the warp factor, which
is to leading order independent of the angular coordinates. The warping
at the tip of the throat is determined by the flux numbers in
(\ref{MK}), 
\begin{equation}
 h_0 \sim \exp \left(-\frac{2\pi K}{3 g_s M}\right). \label{h0}
\end{equation}
The fluxes also stably deform the conifold, therefore setting an IR
cutoff for the radial coordinate~$r$, 
\begin{equation}
  r_0 \sim (g_s M \alpha')^{1/2} h_0. \label{r0}
\end{equation}
Hence the $S^3$ at the throat tip has radius of order $r_0/h_0$. 

Away from the tip, the warp factor is approximated by $h(r) \sim r/R$,
where the curvature radius of $AdS_5$ space is given by 
\begin{equation}
  R^4 \sim g_s N \alpha'^2. \label{R}
\end{equation}

We assume that the bulk CY dominates the six-dimensional volume of the
internal space, i.e. the typical length scale of the internal bulk $L$ 
satisfies $L \gtrsim R$. Hence the four-dimensional Planck mass is 
\begin{equation}
 M_{\mathrm{Pl}}^2 \sim \frac{L^6}{g_s^2 \alpha'^4 }. \label{Planck}
\end{equation}
\\

Before proceeding, let us pause for a moment to explain why we consider
an \aD3-brane and not a D3-brane for the curvaton. This is due to the D3
being free from force in a KS throat, which can be seen as follows: The
form of the self-dual five-form flux is determined by the Bianchi
identity, 
\begin{equation}
 \tilde{F}_5 = (1+*) d\alpha(r, \theta^m) \wedge \sqrt{-g^{(4)}}
  dx^0\wedge dx^1   \wedge dx^2 \wedge dx^3,
 \label{F5}
\end{equation}
where $\alpha$ is a function of the internal coordinates. Then from the
Dirac-Born-Infeld (DBI) and the Chern-Simons (CS) terms in the D-brane
action (cf.~(\ref{DBI-CS})), one can see that a probe D3 feels a potential
proportional to $h^4\! -\! \alpha$, whereas $h^4\! +\! \alpha$ for an
\aD3. Since
\begin{equation}
  h^4 = \alpha \label{halpha}
\end{equation}
is satisfied in a KS throat (and more generally in GKP-type
compactifications with imaginary self-dual fluxes), D3-branes are free
from force. As long as the compactification of the bulk CY is of the
GKP-type, the no-force condition for a D3-brane is preserved, i.e.,
isometry-breaking bulk effects do not give any potential for a D3,
unless the bulk effects drastically distort the throat geometry from the
KS solution. (Note that, by contrast, nonperturbative effects for moduli
stabilization correct (\ref{halpha}) and give similar forces to a D3 and
an \aD3~\cite{DeWolfe:2007hd}. Also, the relation~(\ref{halpha}) is
violated by \aD3-branes.) Since the interplay of bulk effects and  
nonperturbative effects plays an important role in our model, we consider
probe \aD3-branes at the tip of throats.

\subsection{Effective Action of the Curvaton}
\label{subsec:action}

As was explained at the end of the previous subsection, we consider
a probe \aD3-brane inside a GKP-type warped throat. Assuming the \aD3 to
stretch out along the external space, the \aD3 is
driven towards the tip of the throat where the warp factor is
minimized~\cite{Kachru:2002gs}.
In this subsection we study the dynamics of an \aD3-brane in the
angular directions at the throat tip. From these angular directions does
the curvaton(s) arise. 

The effective action for the \aD3's angular degrees of freedom is given
by the DBI and the CS terms, 
\begin{equation}
 S = -T_3 \int d^4 \xi \sqrt{-\det G_{\mu\nu}} 
\left(1- \bar{\Psi} i D \!\!\!\!\slash \, \Psi \right) - T_3 \int C_4,
 \label{DBI-CS}
\end{equation}
where $T_3 \sim 1/(g_s \alpha'^2)$ is the D3 (\aD3) tension. 
Note that in the DBI action we have included the leading order term for
the world-volume fermions~$\Psi$~\cite{Marolf:2003ye,Marolf:2003vf},
into which the angular degrees of freedom (i.e. the curvaton) decay
first. (We assume that (some of) the world-volume fermions are
significantly lighter than the curvaton.) We dropped gauge fields
on the brane since their interactions 
with the curvaton show up from four-point interaction terms and can be
neglected.\footnote{One can check that the
decay of the curvaton into world-volume fermions induced by the
three-point term in (\ref{L}) is the most important
interaction. The curvaton also interacts with world-volume gauge
fields, gravitons, and Kaluza-Klein modes in the throat, but those
interactions have no major effect on the curvaton dynamics,
perturbatively nor through parametric
resonance~\cite{Kofman:1994rk,Kofman:1997yn}.}

When the \aD3 is restricted to the throat tip, the determinant of the
induced metric is
\begin{equation}
\begin{split}
 \det G_{\mu\nu}  &= 
 \det\left( h_0^2 g^{(4)}_{\mu\nu} + h_0^{-2} r_0^2 g^{(3)}_{mn}
 \partial_{\mu} \theta^{m} \partial_{\nu}\theta^n \right) \\
  &\simeq  h_0^8 g^{(4)}
 \left( 1+h_0^{-4} r_0^2 g^{(3)}_{mn} g^{(4)\mu\nu}
 \partial_{\mu}\theta^m \partial_{\nu} \theta^n \right),
 \label{detG}
\end{split}
\end{equation}
where we have set the world-volume coordinates to coincide with the
external coordinates, and $g^{(3)}_{mn}$ denotes the metric of the $S^3$
at the tip. Then by using (\ref{F5}), the four-dimensional Lagrangian
from (\ref{DBI-CS}) is 
\begin{equation}
 \frac{\mathcal{L}}{\sqrt{-g^{(4)}}}\simeq 
 -T_3 (h_0^4 + \alpha_0) - T_3 r_0^2 (\partial \theta)^2 + T_3 h_0^4
 \bar{\Psi} i D \!\!\!\!\slash \, \Psi ,
 \label{3.12}
\end{equation}
where we dropped numerical factors and used the abbreviation:
$ (\partial \theta)^2 \equiv g^{(3)}_{mn} g^{(4)\mu\nu} \partial_{\mu}
\theta^m \partial_{\nu}\theta^n$. We have also ignored terms which
includes more than two derivatives in deriving (\ref{detG}) and
(\ref{3.12}). For simplicity, in most part of this paper we drop
numerical factors and carry out order-of-magnitude estimates. 

So far, the Lagrangian does not include any angular dependent potential
and the \aD3 enjoys the isometry. However, as was briefly mentioned at
the beginning of this section, bulk effects break the throat's
isometries and perturb the background geometry in the following form,
\begin{equation}
 \delta (h^4) \sim h^{\Delta} f(\theta^m),\qquad  
 \delta \alpha \sim h^{\Delta} f(\theta^m). \label{be}
\end{equation}
Note that from the discussion at the end of Subsection~\ref{subsec:WDC},
the perturbation for $h^4$ and $\alpha$ are correlated, especially, both
have the same angular dependence~$f(\theta^m)$. Such effects can be
analysed via gauge/gravity duality, then the power~$\Delta$ of the warp
factor suppressing the perturbations in (\ref{be}) corresponds to the
dimension of the irrelevant operator deforming the dual gauge
theory~\cite{Aharony:2005ez} (see also
\cite{Ceresole:1999zs,Ceresole:1999ht,DeWolfe:2004qx}\footnote{Bulk
effects were also studied in \cite{Baumann:2008kq} in order to investigate
possible corrections to the radial potential of D3-branes for
D-brane inflation. There, general cases beyond GKP-type
compactifications were studied.}). In the KS
throat, the leading perturbation has $\Delta = \sqrt{28} \approx
5.3$, giving a mass (cf. (\ref{mbulk}))
\begin{equation}
 m_{\mathrm{bulk}}^2 \sim \frac{h_0^{3.3}}{g_s M \alpha'} \label{3.29}
\end{equation}
to the canonically normalized angular directions. 

Nonperturbative effects which stabilize the K\"ahler moduli also give
potentials in the angular directions. In \cite{DeWolfe:2007hd},
explicit examples of K\"ahler moduli stabilization due to D7-branes
wrapping four-cycles of conifold throats were studied. For example, in
the simple case of the Kuperstein embedding~\cite{Kuperstein:2004hy} of
the D7-brane, then under the assumption that an \aD3 at the throat tip
lifts the stabilized vacuum energy to a metastable $dS$, the angular
directions receive mass of order 
\begin{equation}
 m_{\mathrm{np}}^2 \sim \frac{h_0^2}{g_s M \alpha'}\frac{\epsilon}{\mu}.
 \label{npmass}
\end{equation}
Here, $\epsilon \sim h_0^{3/2} (g_s M \alpha')^{3/4}$ is the deformation 
parameter of the conifold, and $\mu$ measures the minimal radial
location reached by the D7. However, it should be noted that
(\ref{npmass}) is not necessarily the typical value for our curvaton
model, where multiple throats (e.g. the inflation throat) may exist and
various uplifting mechanisms are possible (see
e.g. \cite{Chen:2008ada}). \\  

Taking into account the isometry breaking bulk effects and
nonperturbative effects, the Lagrangian~(\ref{3.12}) is corrected to
\begin{multline}
 \frac{\mathcal{L}}{\sqrt{-g^{(4)}}}\simeq  -T_3 h_0^4
 - T_3 r_0^2 (\partial \theta)^2 + T_3 h_0^4 \bar{\Psi} i D
 \!\!\!\!\slash \, \Psi \\
 -T_3 h_0^{\Delta} f(\theta^m) - \widetilde{V}_{\mathrm{np}}(\theta^m)
 + T_3 h_0^{\Delta} f(\theta^m) \bar{\Psi} i D \!\!\!\!\slash \, \Psi,
 \label{L1} 
\end{multline}
where we dropped numerical factors of order unity. 
Here $\Delta$ indicates the suppression of the bulk effect by the warp
factor (see (\ref{be})), and we have
represented the angular potential from nonperturbative effects by
$\widetilde{V}_{\mathrm{np}}(\theta^m)$. When there are further effects
which give potentials to the angular position (e.g., the
violation of (\ref{halpha}) due to \aD3-branes), one can formally
include them in $\widetilde{V}_{\mathrm{np}}(\theta^m)$. The
important point here is that $f(\theta^m)$ and
$\widetilde{V}_{\mathrm{np}}(\theta^m)$ generally have different angular
dependence. 

Now we see that various effects induce a gradual but rather bumpy
potential for an \aD3-brane at the throat tip. Among the three angular
directions of the $S^3$ tip, henceforth we focus on the flattest one,
and assume that the \aD3 is stabilized in the other directions. We refer
to this flattest direction as $\theta$, and set the (local) minimum of
$\widetilde{V}_{\mathrm{np}}(\theta)$ as its origin. 
Also, we consider the region where the bulk and nonperturbative effects
are well approximated by expanding the Lagrangian up to quadratic order
in $\theta$, hence (\ref{L1}) can be rewritten as
\begin{multline}
 \frac{\mathcal{L}}{\sqrt{-g^{(4)}}} \simeq 
 -T_3 r_0^2 (\partial \theta)^2 + T_3 h_0^4 \bar{\Psi} i D
 \!\!\!\!\slash \, \Psi  \\
 -m_{\mathrm{bulk}}'^2 (\theta + \theta^2) - m_{\mathrm{np}}'^2 \theta^2
 + m_{\mathrm{bulk}}'^2 (\theta + \theta^2) \bar{\Psi} i D
 \!\!\!\!\slash \, \Psi, 
 \label{L2}
\end{multline}
where $m_{\mathrm{bulk}}'^2 \equiv T_3 h_0^{\Delta}$, 
$m_{\mathrm{np}}'^2 \equiv \partial_{\theta}
\partial_{\theta}\widetilde{V}_{\mathrm{np}}(0)$.
Upon expanding the bulk effects, we neglected numerical factors and
wrote down schematically $f(\theta) \sim \theta^0 + \theta +
\theta^2$, which suffices for our order-of-magnitude estimations. 
Furthermore, we have omitted constant terms which are irrelevant for us.
It should also be noted that as long as $\Delta >4$, terms of the form
$T_3 h_0^{\Delta} \bar{\Psi} i D \!\!\!\!\slash \, \Psi$ can well be
ignored.  

Let us again reparameterize $\theta$ so that the minimum of its total
potential (i.e. the first two terms in the second line of (\ref{L2}))
becomes the origin. The shift is roughly
\begin{equation}
 \theta \longrightarrow \theta - \frac{m_{\mathrm{bulk}}'^2}{2
  (m_{\mathrm{np}}'^2 + m_{\mathrm{np}}'^2)}.
\end{equation}
Canonically normalizing the fields,
\begin{equation}
 \sigma \equiv \sqrt{T_3} r_0 \theta, \qquad
 \psi \equiv \sqrt{T_3} h_0^2 \Psi ,
 \label{canor}
\end{equation}
we arrive at
\begin{equation}
 \frac{\mathcal{L}}{\sqrt{-g^{(4)}}} \simeq -(\partial \sigma)^2 +
  \bar{\psi}i D \!\!\!\!\slash \, \psi - (m_{\mathrm{bulk}}^2 +
  m_{\mathrm{np}}^2) \sigma^2 + \frac{g_s M^{1/2} \alpha'^{3/2}}{h_0^3}
  \frac{m_{\mathrm{bulk}}^2
  m_{\mathrm{np}}^2}{m_{\mathrm{bulk}}^2+m_{\mathrm{np}}^2}\sigma 
  \bar{\psi}i D \!\!\!\!\slash \, \psi,
 \label{L}
\end{equation}
where we have neglected four-point and higher interaction terms. The
masses arising from bulk and nonperturbative effects are defined as
follows, respectively,
\begin{equation}
 m_{\mathrm{bulk}}^2 \equiv \frac{m_{\mathrm{bulk}}'^2}{T_3 r_0^2} =
  \frac{h_0^{\Delta}}{r_0^2} \sim \frac{h_0^{\Delta -2}}{g_s M \alpha'}
  , \label{mbulk}
\end{equation}
\begin{equation}
 m_{\mathrm{np}}^2 \equiv \frac{m_{\mathrm{np}}'^2}{T_3 r_0^2 } \equiv 
  \frac{h_0^{\lambda-2}}{g_s M \alpha'} . \label{mnp}
\end{equation}
In the far right hand side of (\ref{mnp}), we have additionally
introduced the symbol~$\lambda$ in order to denote the
suppression of the nonperturbative effects (or more generally, the sum
of all effects except from the bulk). By comparing $\Delta $ and
$\lambda$, one can measure the relative strength of the bulk and
nonperturbative effects at the throat tip, e.g., in the specific case of
(\ref{3.29}) and (\ref{npmass}) with $\epsilon / \mu \sim h_0^{3/2}$,
then $\Delta \sim 5.3$ and $\lambda \sim 5.5$. 
The point we would like to emphasize is that various effects with
different angular dependence misalign the (local) minima of the
potential and interaction terms, hence providing decay channels to the
curvaton. 

Let us summarize what we have obtained. The curvaton candidate~$\sigma$
has mass
\begin{equation}
 m^2 \sim m_{\mathrm{bulk}}^2 + m_{\mathrm{np}}^2,
\end{equation}
and its decay rate into world-volume fermions is
\begin{equation}
 \Gamma \sim
 \left(\frac{g_s M^{1/2}\alpha'^{3/2}}{h_0^3} \frac{m_{\mathrm{bulk}}^2
 m_{\mathrm{np}}^2}{m_{\mathrm{bulk}}^2+m_{\mathrm{np}}^2}\right)^2 m^3
 \sim \frac{g_s^2 M \alpha'^3}{h_0^6} 
 \frac{m_{\mathrm{bulk}}^4 m_{\mathrm{np}}^4}{(m_{\mathrm{bulk}}^2 +
 m_{\mathrm{np}}^2)^{1/2}}. \label{gamma}
\end{equation}
Specifically, when bulk effects are dominant over nonperturbative effects,
i.e. $\Delta < \lambda$, then 
\begin{equation}
 m^2 \sim \frac{h_0^{\Delta-2}}{g_s M \alpha'}, \qquad
 \Gamma \sim \frac{h_0^{2 \lambda + \frac{3}{2} \Delta -13}}{g_s^{3/2}
 M^{5/2} \alpha'^{1/2}}. \label{bulkdom}
\end{equation}
The opposite case where nonperturbative effects are stronger can be
treated simply by exchanging $\Delta$ and $\lambda$. 

The field range of $\sigma$ is restricted by the radius of the $S^3$
tip. From (\ref{canor}),
\begin{equation}
 \sigma \lesssim \sqrt{T_3} r_0 \sim \frac{h_0 M^{1/2}}{\alpha'^{1/2}}. 
\end{equation}
Then statistically one expects that the field value of the curvaton
during inflation is
\begin{equation}
 \sigma_* \sim \frac{h_0 M^{1/2}}{\alpha'^{1/2}}.  \label{sigmastar}
\end{equation}
We assume that the number of peaks and valleys of the periodic angular
potential to be of order one, so that the approximation of the
quadratic potential in (\ref{L}) be valid. Here, from (\ref{R}) and
(\ref{Planck}) one finds that the condition~(\ref{3}) is always
satisfied under $R \lesssim L$ and the weak string coupling~$g_s < 1 $,
\begin{equation}
 \frac{\sigma_*}{M_{\mathrm{Pl}}} \sim \frac{h_0 g_s^{1/4}}{N^{1/4} K^{1/2}}
  \left(\frac{R}{L}\right)^3 <1.
 \label{trivial3}
\end{equation}

\subsection{Consistency Conditions}
\label{subsec:con}

Before moving on to discuss cosmology, we should mention some consistency
conditions for the above analyses. In addition to the requirements
discussed in Section~\ref{sec:curvaton}, the following microscopic
constraints will further restrict the parameter space. 

\subsubsection*{Speed Limit of the Curvaton}

We have expanded the DBI action up to two derivatives, hence for the
validness of this procedure the curvaton has to be moving
nonrelativistically. This requirement is also important for avoiding
significant back reaction of the curvaton brane to the throat. 

The light speed at the tip is given by $\dot{\theta}_{\mathrm{rel}} \sim
h_0^2/r_0$, whereas the maximal speed of a harmonically oscillating
curvaton is $\dot{\theta}_{\mathrm{max}} \sim m$. Therefore we require 
\begin{equation}
 \frac{mr_0}{h_0^2} \sim \frac{m (g_s M \alpha')^{1/2}}{h_0}< 1.
  \label{a} 
\end{equation}
From (\ref{mbulk}) and (\ref{mnp}) it is clear that as long as $\Delta,
\lambda > 4$, the curvaton brane is nonrelativistic. 

\subsubsection*{Stringy Corrections to the Throat}

One should be aware of stringy corrections during inflation. When the
Hubble parameter during inflation is larger than the local string scale
of the throat, such corrections can become significant and may even
shorten the throat~\cite{Frey:2005jk}, therefore we require
\begin{equation}
 H_{\mathrm{inf}} \frac{\alpha'^{1/2}}{h_0} < 1. \label{b}
\end{equation}
Note that this condition can be rewritten in terms of
$\sigma_*$~(\ref{sigmastar}) as 
\begin{equation}
 M^{1/2} \frac{H_{\mathrm{inf}}}{\sigma_*} < 1, \label{bprime}
\end{equation}
which guarantees the density perturbation from the curvaton to be nearly
Gaussian~(\ref{6}).\footnote{Our observable universe might have happened
to be in a place which is less likely than the statistically preferred
$\sigma_*$~(\ref{sigmastar}), e.g., extremely near the (local)
minimum/maximum of the periodic angular potential. In such case,
(\ref{6}) can be violated and the curvaton may contribute to producing
large non-Gaussianity.} 

\subsubsection*{Curvaton's Energy Scale and the Local String Scale}

The curvaton's oscillation energy also should be smaller than the local
string scale for avoiding serious stringy corrections in the throat,
hence 
\begin{equation}
 m^2 \sigma_*^2 \left( \frac{\alpha'^{1/2}}{h_0}\right)^4
 \sim \frac{m^2 M \alpha'}{h_0^2} < 1.
 \label{c}
\end{equation}
One sees that this requirement contains the speed limit
condition~(\ref{a}) provided $g_s < 1$. \\

Before ending this section, let us mention that the condition
\begin{equation}
 \Delta , \lambda> 4 \label{Dl4}
\end{equation}
following from (\ref{a}) and (\ref{c}) can also be understood as the
requirement that the bulk and nonperturbative effects be weak
corrections to the original angular independent background. (One way of
understanding this is by looking at the induced mass (\ref{mbulk}) and
(\ref{mnp}): One expects from naive power counting that $m^2 \propto
h_0^2$, whereas the actual mass is $m^2 \propto h_0^{\Delta-2},
h_0^{\lambda-2}$.) 
We also note that under (\ref{Dl4}), the
condition~(\ref{4}) becomes trivial. It suffices to show this in the
bulk effect dominant case~(\ref{bulkdom}), where 
\begin{equation}
 \frac{\Gamma}{m} \sim \frac{h_0^{2 \lambda + \Delta -12}}{g_s M^2} < 1.
 \label{trivial4}
\end{equation}

\section{Generating the Cosmological Perturbations}
\label{sec:perturbations}

Equipped with the information on the angular position of the \aD3-brane
and the consistency conditions discussed in the previous sections, let us
look into the parameter space and show that the angular oscillation of
the \aD3 at the throat tip actually plays the role of the curvaton. In
this section we investigate several scenarios of interest. First we
consider the case where the curvaton generates the observed curvature
perturbation (and in some situations large non-Gaussianity) in
Subsection~\ref{subsec:cobe}. Then, 
cases where the curvaton contributes in generating large non-Gaussianity
either through curvature or isocurvature perturbations are discussed,
respectively, in Subsection~\ref{subsec:nG-adi} and
\ref{subsec:nG-iso}. 

We assume that the SM particles are realized on the curvaton \aD3-brane
in Subsection~\ref{subsec:cobe} and \ref{subsec:nG-adi}. How the SM
particles are realized on the world-volume of 
the D-brane is out of the scope of this paper, though one naively
expects that the interactions among the open string modes on the brane
is suppressed by the local string scale. As can be seen from
(\ref{gamma}), or more explicitly from (\ref{bulkdom}) and (\ref{Dl4}),
the decay rate from the curvaton to the world-volume fermions is further 
suppressed by the warp factor. Therefore, after the
curvaton decay into world-volume fermions, we can expect the
fermions to soon decay into or thermalize with the SM particles (or
particles that eventually turn into SM particles). 
In Subsection~\ref{subsec:nG-iso}, the SM brane can be far apart from
the curvaton brane.

For the cosmic inflation, we do not specify its details. However,
throughout this paper we have assumed that the inflaton energy is
transferred to radiation right after the end of inflation. For instance,
one can have in mind brane-antibrane inflation in a warped throat
(possibly, a throat different from the throat containing the curvaton
brane). From the brane-antibrane annihilation which ends inflation,
heavy closed strings are produced, which further decay into lighter
states. Such processes happen right after inflation ends, and if there
are branes left, the inflation energy will soon be converted to
radiation on the world-volume of the residual brane. (However, 
one may have to be careful with stable angular KK modes, see
e.g. \cite{Chen:2006ni,Dufaux:2008br}. Of course, such problems do not
arise in other inflation mechanisms, such as modular inflation models.) 

We should also comment on the ``overshooting problem'' which states that
the Hubble parameter during inflation is bounded by the gravitino mass,
i.e. $H_{\mathrm{inf}} \lesssim m_{3/2}$~\cite{Kallosh:2004yh}, in the
simplest inflationary models based on the moduli stabilization scenario
of \cite{Kachru:2003aw}. Requiring a gravitino mass of
$\mathcal{O}(1) \mathrm{TeV}$, the inflation energy scale is restricted
to be very low. If the curvaton generates the observed
curvature perturbations, i.e. the inequality (\ref{7.5}) being
saturated, then one can show from the requirements laid out in
Section~\ref{sec:curvaton} that the Hubble parameter during inflation
cannot be that small. However, as we briefly mentioned in
Section~\ref{sec:intro}, recent studies have demonstrated several
ways to cure this problem, e.g., by introducing more general class of
K\"ahler moduli stabilization with superpotential of the racetrack form,
with positive exponents, and so on
\cite{Kallosh:2004yh,Badziak:2008yg,Abe:2008xu,Conlon:2008cj,Badziak:2008gv,Chen:2009nk}.
We expect such mechanisms to be operating and the inflation energy scale
be disentangled from the gravitino mass.

\subsection{Curvaton Scenario}
\label{subsec:cobe}

Let us consider the case where the angular fluctuation of the 
\aD3-brane at the throat tip generates the observed primordial curvature
perturbation. We assume the SM particles to be realized on the curvaton
\aD3-brane. 

In cases where the curvaton dominates the universe before
it decays, the reheating of the universe is sourced by the decay of the
curvaton. We require the curvaton to decay before BBN, but if one
also wants to incorporate baryogenesis, then the
condition~(\ref{5}) should be corrected according to the baryogenesis
scenario. 

On the other hand, if the curvaton is subdominant at decay,
then reheating should rely on the remnants of inflation, e.g., one
can imagine the source that drives inflation to be also in the SM throat
(i.e. the curvaton throat) and that the inflation energy directly flows
into the SM brane, or it could also be that the inflation energy tunnels
into the SM throat from the bulk or from some other throat
\cite{Chen:2006ni,Dimopoulos:2001ui,Dimopoulos:2001qd,Barnaby:2004gg,Kofman:2005yz,Chialva:2005zy,Langfelder:2006vd,Harling:2007jy}. Although
we keep our discussion general and do not go into details beyond the
curvaton model, let us address some issues which may rise in the 
curvaton-subdominant case: Depending on how the inflation energy is
transported to the SM brane, curvaton production after inflation may
happen and the curvaton contribution to the curvature perturbation may be
reduced~\cite{Linde:2005yw}. Also, large isocurvature perturbation in
cold dark matter (CDM) or baryons can be produced if such particles are
created significantly before the curvaton decay, and/or if the curvaton
itself creates
them~\cite{Lyth:2002my,Lyth:2003ip,Kawasaki:2008sn,Langlois:2008vk,Moroi:2008nn,Kawasaki:2008pa}. \\ 

Recall from the discussions in Section~\ref{sec:curvaton} and
Section~\ref{sec:CWT} that after dropping the conditions that trivially
follow from the other ones, we arrive at the following six conditions:
(\ref{2}), (\ref{5}), (\ref{7.5}), (\ref{8}), (\ref{b}),
(\ref{c}). Here, especially, (\ref{7.5}) should be saturated since we
require the curvaton to generate the observed curvature
perturbation. The $f_{\mathrm{NL}}$ constraint is now simply
\begin{equation}
 f_{\mathrm{NL}} \sim \frac{1}{r} \lesssim 100, \label{4.1}
\end{equation}
and the inflation energy scale and the produced non-Gaussianity are
related by 
\begin{equation}
 \frac{H_{\mathrm{inf}}}{M_{\mathrm{Pl}}} \sim 10^{-5} h_0 g_s M^{1/2}
  \left( \frac{\alpha'^{1/2}}{L}\right)^3 f_{\mathrm{NL}}.
\end{equation}
Under (\ref{7.5}) and (\ref{4.1}), the condition~(\ref{b}) is
automatically satisfied,
\begin{equation}
 H_{\mathrm{inf}} \frac{\alpha'^{1/2}}{h_0} \lesssim
  \frac{M^{1/2}}{10^3} < 1,
\end{equation}
where we have used the known maximal value of the CY four-fold Euler
number~(\ref{euler}) in obtaining the second inequality.

Therefore, after fixing the inflation energy scale from the saturated
(\ref{7.5}), the independent conditions that restrict the parameter space
are (\ref{2}), (\ref{5}), (\ref{8}), and (\ref{c}). Especially in the
bulk effect dominant case (i.e. $\Delta < \lambda$)~(\ref{bulkdom}), the
four conditions take the following form, respectively,
\begin{equation}
 r^2 \frac{10^{10} h_0^{\Delta-4}}{g_s M^2} \ll 1, \qquad
 \frac{10^{42} h_0^{2 \lambda + \frac{3}{2} \Delta -13}}{g_s^{1/2}
  M^{5/2} } \left(\frac{\alpha'^{1/2} }{L}\right) ^3 > 1, \qquad
 \frac{1}{r} \lesssim 10^2, \qquad
  \frac{h_0^{\Delta-4}}{g_s} < 1,
\end{equation}
where 
\begin{equation}
 r \sim \mathrm{min}
 \left\{ h_0^{8-\lambda - \frac{\Delta}{2}} g_s^{5/2} M^2 
 \left(\frac{\alpha'^{1/2}}{L}\right)^6, \, 1 \right\}.
\end{equation}

We illustrate the parameter space on the $\Delta$-$\lambda$ plane in
Figure~1\footnote{As we carry out only an order-of-magnitude 
estimate, the explicit values of the parameters for each figure should
not be taken so seriously, e.g., the actual $L/\alpha'^{1/2}$ is larger
than the set values since we dropped numerical factors in
(\ref{Planck}).}. One sees that as $\Delta$ or $\lambda$ increase, 
i.e. the bulk or nonperturbative effects weaken, the decay rate of the
curvaton is suppressed and the curvaton comes closer to dominating the
universe before decay. Also, the produced non-Gaussianity (displayed in
the figure as contour lines) decreases as $r \to 1$. The figure
clearly shows that when either the bulk or nonperturbative effect is
absent, i.e. $\Delta$ or $\lambda$ $\to \infty$, the curvaton cannot
decay,\footnote{Strictly speaking, the curvaton does decay due to the
interaction terms we have ignored in deriving (\ref{L}), but such
effects are negligibly small.} and one crosses the orange
line which is the BBN constraint~(\ref{5}).  

As we take different values for the four parameters $g_s$, $M$,
$L/\alpha'^{1/2}$, and $h_0$, the consistent region (the yellow region
in the figure) deforms and shifts in the $\Delta$-$\lambda$
plane. It can also split into $r \gtrless 1$ regions when $h_0$ becomes
large. Though we have shown only a single example, one can check that
consistent curvaton scenarios are allowed for broad ranges of the
parameters.  

\begin{figure}[htb]
\begin{center}
\includegraphics[width=0.5\linewidth]{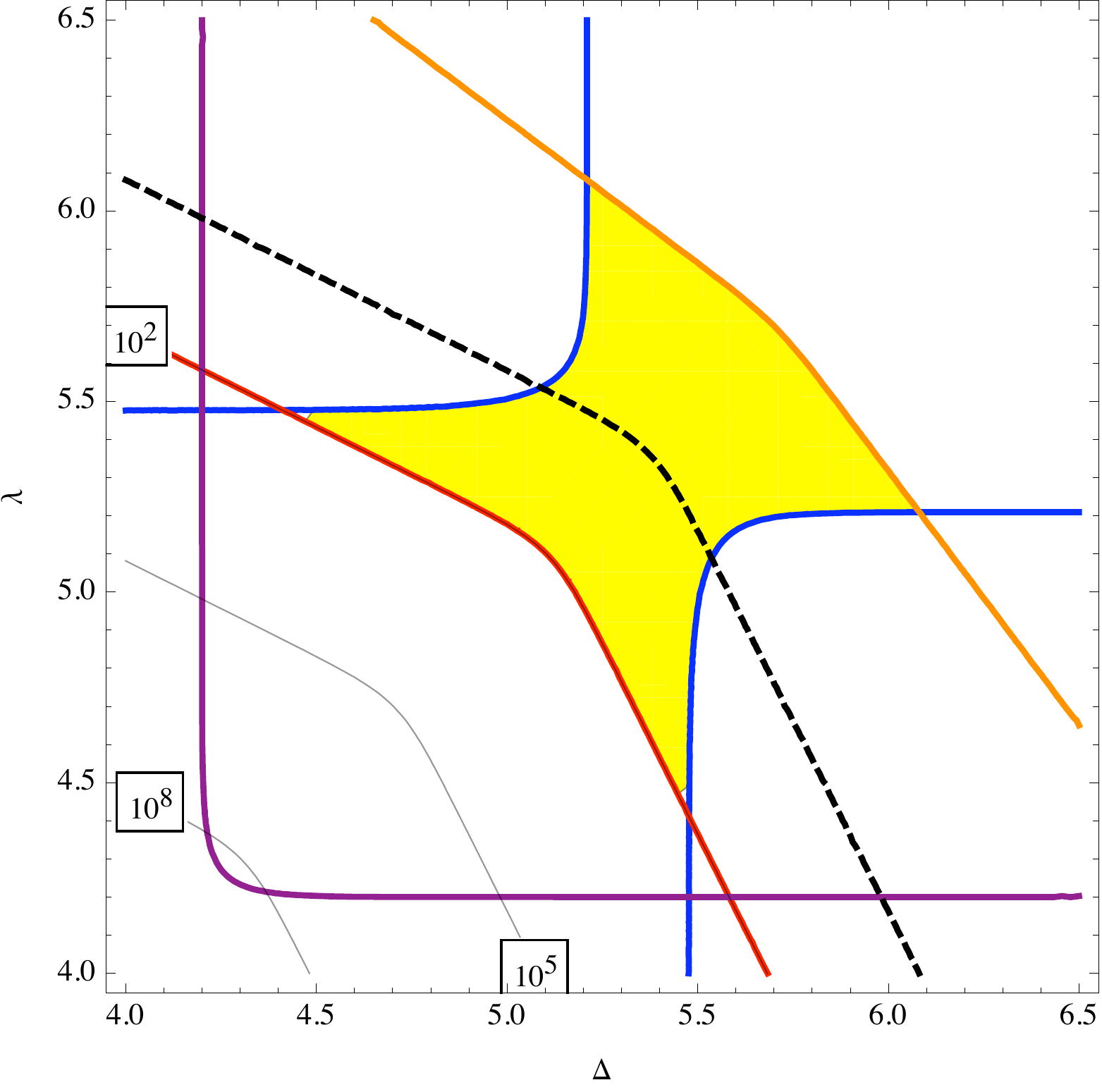}
\end{center}
 {\bf Figure 1:}
 The consistency conditions for the curvaton scenario on the
 $\Delta$-$\lambda$ plane, where the  other parameters are set
 to $g_s = 0.1$, $M=300$, $L/\alpha'^{1/2} = 3$, $h_0=10^{-5}$ (hence $N
 \sim 50000$, $K \sim 200$, $M_{\mathrm{Pl}} \alpha'^{1/2} \sim
 300$). The lines denote where each condition is saturated, blue:
 masslessness~(\ref{2}), orange: BBN~(\ref{5}), red:
 $f_{\mathrm{NL}}$~(\ref{8}), purple: curvaton oscillation
 energy~(\ref{c}). The yellow region satisfies all four conditions. On
 the right (left) side of the dashed line, the curvaton dominates
 (subdominates) the universe at the decay epoch. The produced
 non-Gaussianities are also shown as contour lines for
 $f_{\mathrm{NL}}$. Here the inflation scale can be estimated
 from $H_{\mathrm{inf}}/M_{\mathrm{Pl}} \sim 10^{-11} f_{\mathrm{NL}}$.
\end{figure}

\subsection{Non-Gaussianity through Curvature Perturbations}
\label{subsec:nG-adi}

Next we consider the case where the curvaton produces
large non-Gaussianity, but generates only negligible curvature
perturbations. Hence in this subsection we assume some other mechanism
(e.g. the inflaton) to be responsible for generating the observed curvature
perturbations. As in the previous subsection, the SM particles are
assumed to be realized on the curvaton \aD3-brane. 

Since we consider curvatons generating large non-Gaussianity, we
focus on cases where the curvaton decays before dominating the
universe. (One clearly sees from (\ref{7.5}) and (\ref{8}) that when
$r\sim 1$, the amplitude constraint is severer than the
$f_{\mathrm{NL}}$ constraint.) If the energy fraction of the curvaton is
sufficiently small, then it can survive beyond the BBN epoch without
ruining nucleosynthesis. 

The independent consistency conditions are (\ref{2}), (\ref{7.5}),
(\ref{8}), (\ref{b}), (\ref{c}), and depending on $r$, the BBN
constraint~(\ref{5}) is also required. In Figure~2 we
show the conditions in the $\Delta$-$\lambda$ plane under a certain set
of parameters. Note that we now have an additional
parameter~$H_{\mathrm{inf}}/M_{\mathrm{Pl}}$ since (\ref{7.5}) need not be
saturated.
On the red line in the figure, the curvaton produces the observationally
allowed maximum non-Gaussianity of order $f_{\mathrm{NL}} \sim 100$. 
In contrast to the previous subsection, the non-Gaussianity increases
for larger $\Delta$ or $\lambda$, since then $r$ becomes larger and the
non-Gaussian curvature perturbations generated by the curvaton increase.

\subsection{Non-Gaussianity through Isocurvature Perturbations}
\label{subsec:nG-iso}

In this subsection, we consider the case where the curvaton survives
until now and contribute to dark matter. Here the SM particles need not
be on the curvaton brane. Specifically, in cases where the
throat containing the curvaton brane is geometrically separated from
where the SM particles are realized, then the curvaton dark matter
interacts with SM particles only through graviton mediation and thus
behaves as hidden dark matter (see e.g. \cite{Chen:2006ni}). 

Since the curvaton does not decay, here we need not require both the
bulk and nonperturbative effects, and one of them can be totally absent,
i.e. $\Delta$ or $\lambda$ $\to \infty$. D3-branes which do
not feel bulk effects are thus always of this type (unless there are
additional effects giving angular dependencies.)

As for the cosmological perturbations, the effects of having such
curvatons are similar to that of axions. Since they are decoupled from
(most of) the thermal history of the SM particles, their fluctuations 
entirely contribute to isocurvature perturbations. Here we discuss
non-Gaussianity produced from the curvaton through isocurvature
perturbations. As in the previous subsection, we assume the universe to
be reheated by the remnants of inflation, and the observed curvature
perturbations to be generated by some mechanism other than the curvaton. 

Here, additional constraints are required. Instead of the BBN
constraint~(\ref{5}), we require the curvaton to be stable until
now, therefore
\begin{equation}
 10^{60} \frac{\Gamma}{M_{\mathrm{Pl}}} < 1. \label{a1}
\end{equation}
Also, the present energy density of the curvaton should not exceed that
of dark matter, $\Omega_{\sigma 0} \lesssim
\Omega_{\mathrm{CDM}0}$. Assuming radiation domination from the end of
inflation to the time of matter-radiation equality, the present curvaton
energy density is $\rho_{\sigma 0} \sim m^2 \sigma_*^2
(t_{\mathrm{osc}}/t_{\mathrm{eq}})^{3/2} (a_{\mathrm{eq}}/a_0)^{3}$,
and one arrives at the requirement
\begin{equation}
 R \equiv \frac{\Omega_{\sigma 0}}{\Omega_{\mathrm{CDM} 0}} \sim 
10^{28} \frac{m^{1/2} \sigma_*^2}{M_{\mathrm{Pl}}^{5/2}} \lesssim 1. \label{a2}
\end{equation}
Fluctuations of the curvaton dark matter give rise to isocurvature
perturbations between CDM and radiation,
\begin{equation}
 \mathcal{S}_{\mathrm{CDM}} \sim R \left[2
				    \frac{\delta\sigma_*}{\sigma_* } + 
			  \left(\frac{\delta
			   \sigma_*}{\sigma_*}\right)^2  \right]. 
\end{equation}
Here the isocurvature and curvature perturbations are expected to be
uncorrelated, hence the observational bound on isocurvature
perturbation~\cite{Komatsu:2008hk} imposes
\begin{equation}
 \frac{\mathcal{P}_{\mathcal{S}}}{\mathcal{P}_{\zeta}} \sim
  \frac{R^2}{\mathcal{P}_{\mathrm{COBE}}} 
 \left(  \frac{H_{\mathrm{inf}}}{\sigma_*}\right)^2 
  \lesssim \frac{1}{10}, \label{a3}
\end{equation}
where $\mathcal{P}_{\mathrm{COBE}} \sim 10^{-10}$. 
Also, the non-Gaussianity from isocurvature perturbation in terms of the
non-linearity parameter is
\begin{equation}
 f_{\mathrm{NL}} \sim \frac{R^3}{\mathcal{P}_{\mathrm{COBE}}^{2}} \left(
			   \frac{H_{\mathrm{inf}}}{\sigma_*}\right)^4 
  \lesssim 100 . \label{a4}
\end{equation}
One should note that the non-Gaussianity from isocurvature and curvature
perturbations manifest themselves differently in the CMB temperature
fluctuations. Especially, non-Gaussian effects from isocurvature
perturbations are enhanced in the large scales~\cite{Kawasaki:2008sn}. 
For detailed analyses on this issue, see
\cite{Kawasaki:2008sn,Langlois:2008vk,Moroi:2008nn,Kawasaki:2008pa,Hikage:2008sk}. \\

Now (\ref{5}), (\ref{7.5}), and (\ref{8}) are replaced by
the above conditions, and we arrive at the following seven independent
consistency conditions: (\ref{2}), (\ref{b}), (\ref{c}), (\ref{a1}),
(\ref{a2}), (\ref{a3}), and (\ref{a4}). In Figure~3 we show an example
where the KS throat with $\Delta = \sqrt{28}$ bulk effect can produce
large non-Gaussianity. In the example, the consistent region is bounded
by the isocurvature bound~(\ref{a3}) rather than the $f_{\mathrm{NL}}$
bound~(\ref{a4}). Larger curvaton mass leads to larger $R$, therefore
larger isocurvature perturbation and non-Gaussianity.  \\

In this subsection we have studied curvatons that do not decay.
However, in cases where the curvaton brane is geometrically
separated from the SM sector, the curvaton can decay and still
generate large non-Gaussianity. The light open string modes excited from
the curvaton decay will serve as hidden dark radiation and/or matter and
can have significant effects through isocurvature perturbations,
as in the case of the non-decaying curvatons. It would be interesting to
examine systematically the range of possibilities arising from
oscillating D-branes in multi-throat compactifications. 

\begin{figure}[htb]
\begin{center}
\begin{minipage}{.48\linewidth}
\includegraphics[width=\linewidth]{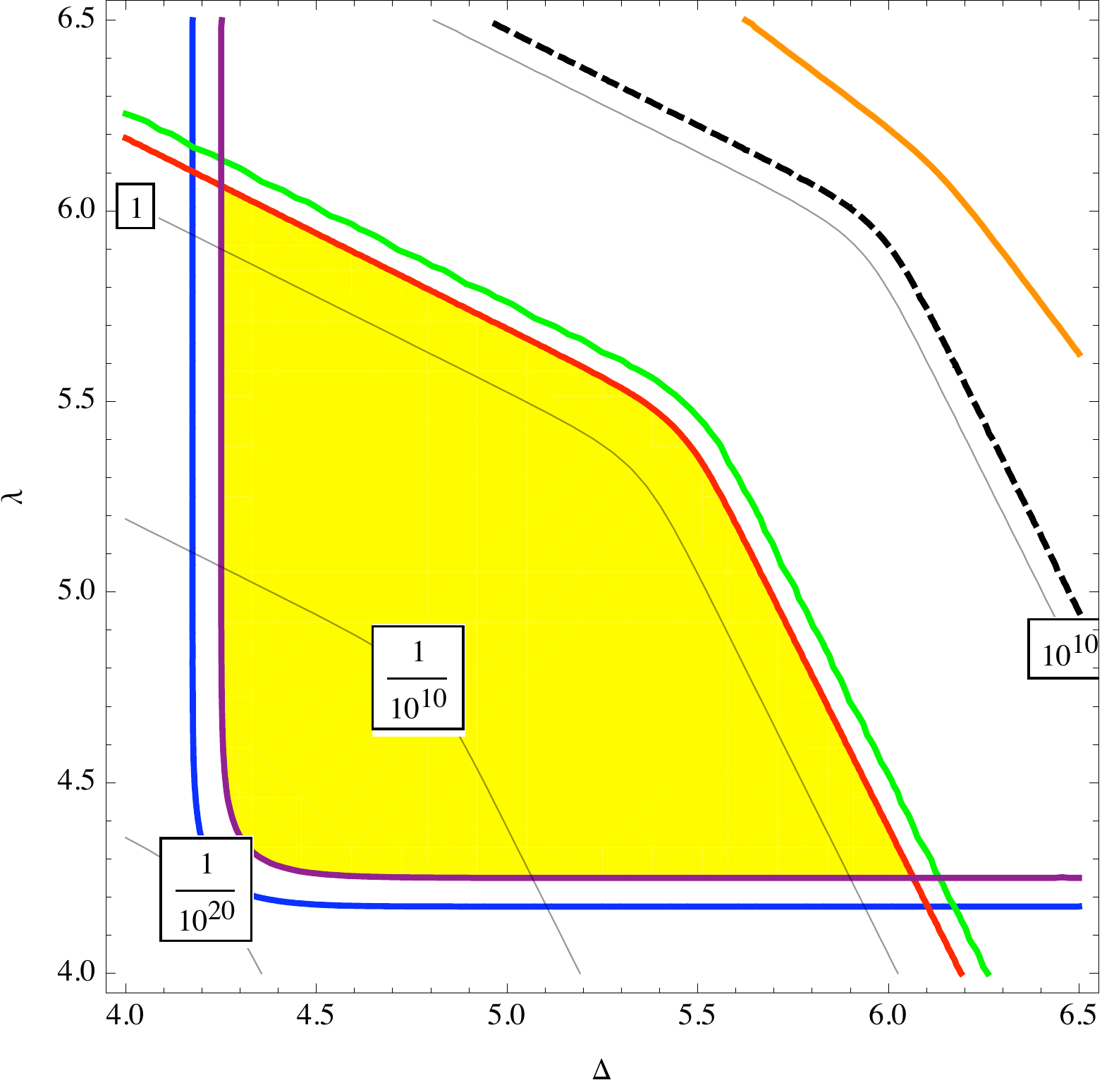}
\end{minipage}
\hspace{0.0pc}
\begin{minipage}{.48\linewidth}
\includegraphics[width=\linewidth]{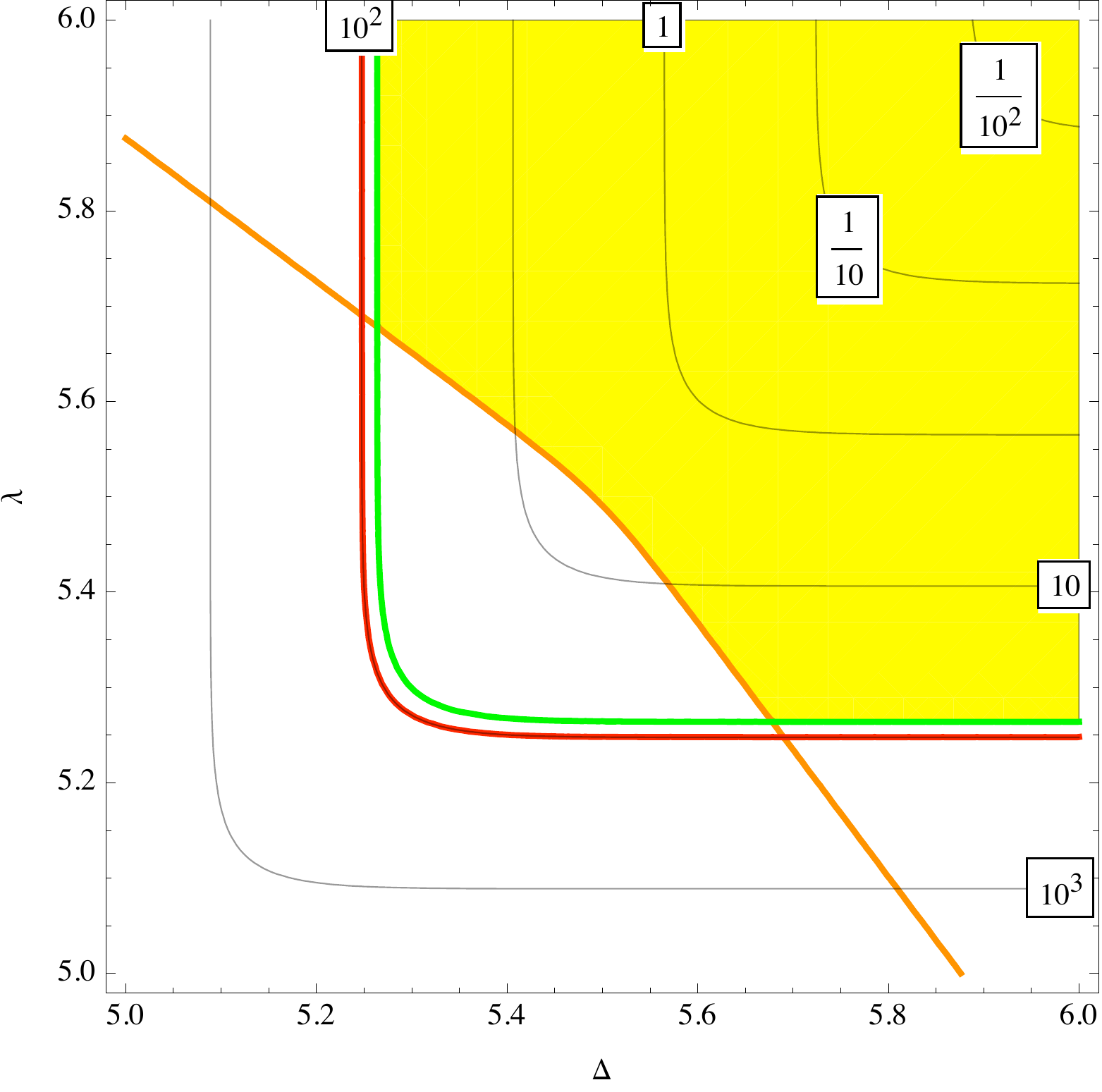}
\end{minipage}
\end{center}
 {\bf Figure 2 (left):}
 The consistency conditions on the $\Delta$-$\lambda$ plane for
 the case where the curvaton contributes mainly to 
 non-Gaussianity. The parameters are set to $g_s = 0.1$, $M=200$,
 $L/\alpha'^{1/2} = 10$, $h_0=10^{-4}$,
 $H_{\mathrm{inf}}/M_{\mathrm{Pl}} = 10^{-9}$ (hence $N \sim 20000$, $K
 \sim 90$, $M_{\mathrm{Pl}} \alpha'^{1/2} \sim 10000$). The 
 lines denote where each condition is saturated, blue:
 masslessness~(\ref{2}), orange: BBN~(\ref{5}), green: curvature 
 perturbation~(\ref{7.5}), red: $f_{\mathrm{NL}}$~(\ref{8}), purple:
 curvaton oscillation energy~(\ref{c}). The stringy correction~(\ref{b})
 condition is satisfied in the entire displayed area. The yellow region
 satisfies all six conditions. Also, the dashed line denotes where the
 curvaton starts to dominate the universe before decay. The produced
 non-Gaussianities are shown as contour lines for
 $f_{\mathrm{NL}}$. \\ 

 {\bf Figure 3 (right):}
 The consistency conditions on the $\Delta$-$\lambda$ plane for 
 the case where the curvaton is stable until now. The parameters are set
 to $g_s = 0.1$, $M=100$,  $L/\alpha'^{1/2} = 10$, $h_0=4 \times
 10^{-9}$, $H_{\mathrm{inf}}/M_{\mathrm{Pl}} = 10^{-13}$ (hence
 $N \sim 9000$, $K \sim 90$, $M_{\mathrm{Pl}} \alpha'^{1/2} \sim 10000$). 
 The lines denote where each condition
 is saturated, orange: stability~(\ref{a1}), green: isocurvature
 perturbation~(\ref{a3}), red: $f_{\mathrm{NL}}$~(\ref{a4}). The
 masslessness~(\ref{2}), stringy correction~(\ref{b}), curvaton
 oscillation energy~(\ref{c}), and DM~(\ref{a2}) conditions are satisfied 
 in the entire displayed area. The yellow region satisfies all seven
 conditions. The produced non-Gaussianities are shown as contour lines
 for  $f_{\mathrm{NL}}$.
\end{figure}

\section{Conclusions}
\label{sec:con}

We have proposed a model for generating the primordial
perturbations and reheating our universe from angular oscillations of
D-branes at the tip of throats. The geometrical features of throats
-- warping and (approximate) isometries -- yielded curvaton scenarios. 
We have also seen that effects that break the force-free
condition of the D-brane in the isometry directions, such as the
isometry breaking bulk effects and moduli stabilizing nonperturbative
effects played an important role in our model. Depending on the
(un)balance between the various features of the background geometry, the
curvaton model showed different behaviours. We have studied cases where 
the curvaton generated the observed curvature
perturbations (Subsection~\ref{subsec:cobe}), the curvaton contributed
mainly to non-Gaussianity through curvature perturbations
(Subsection~\ref{subsec:nG-adi}), and through isocurvature perturbations
(Subsection~\ref{subsec:nG-iso}). Each scenario could be realized in a
wide range of parameter space. In other words, our model may be
considered as generally arising from compactification scenarios
containing warped throats with isometries. Especially the case
considered in Subsection~\ref{subsec:nG-iso} arise no matter where the
SM particles are realized, and therefore may serve as a test for
discussing the validness of (multi-)throat compactification
scenarios. (See also \cite{Dufaux:2008br} for discussions in this
direction.) The curvaton model is capable of producing large
non-Gaussianity, and upcoming CMB experiments are expected to allow us
to give more rigorous arguments.  

In this paper we tried to keep our discussions general and did not go
into details of individual moduli stabilization mechanisms. We have
treated the strengths of the bulk and nonperturbative effects as free
parameters $\Delta$ and $\lambda$, and only made
order-of-magnitude estimations. However, based on previous studies,
one can carry out more elaborate analyses on the angular potentials for
D-branes at the $S^3$ tip of warped deformed conifolds, for certain
moduli stabilizing scenarios. It is necessary and interesting to
explore the throat tip potential and also a wide class of
compactification scenarios for concrete realizations of the curvaton. 
We note again that in most part of the paper we simply assumed the SM
particles to be realized on the curvaton \aD3-brane(s). Inclusion of a
more realistic SM sector is necessary for a fully consistent model, and
is crucial for detailed studies of the thermal history of the universe. 

One of the general lessons of our work is that in string theory, one
finds it quite natural to consider fields other than the inflaton for
generating the primordial perturbations. Such light fields can be
realized in string theory, thus giving rise to mechanisms which may have
seemed too intricate from the phenomenological point of view. It is fair
to say that top-down approaches to inflationary cosmology can provide us
with rich ideas beyond the standard slow-roll inflation pictures.

\section*{Acknowledgements}

We would like to thank Robert Brandenberger, Damien Easson, Toshiya
Imoto, Etsuko Kawakami, Shunichiro Kinoshita, Kazunori Nakayama, Brian
Powell, Misao Sasaki, Toyokazu Sekiguchi, Gary Shiu, Shigeki Sugimoto,
Fuminobu Takahashi, Atsushi Taruya, Henry Tye, Yi Wang, and Jun'ichi
Yokoyama for very helpful discussions. T.K. is also grateful to Masahiro
Kawasaki and Katsuhiko Sato for their continuous support. 
The work of S.M. was supported in part by MEXT through a Grant-in-Aid
for Young Scientists (B) No.~17740134, by JSPS through a
Grant-in-Aid for Creative Scientific Research No.~19GS0219, and by the
Mitsubishi Foundation. This work was supported by World Premier
International Research Center Initiative (WPI Initiative), MEXT, Japan.


\end{document}